\documentclass[english,aps,prd,reprint,showpacs,floatfix,nofootinbib,twocolumn]{revtex4}

\usepackage{color}
\usepackage{babel}
\usepackage{amsmath}
\usepackage{amssymb}
\usepackage{graphicx}
\usepackage{longtable}
\usepackage{dcolumn}

\def\a{\alpha}
\def\r{\rho}
\def\s{\sigma}
\def\t{\tau}
\def\m{\mu}
\def\n{\nu}
\def\k{\kappa}
\def\th{\theta}
\def\g{\gamma}\def\G{\Gamma}
\def\L{\Lambda}\def\l{\lambda}
\def\D{\Delta}
\def\la{\langle}
\def\ra{\rangle}
\def\o{\omega}\def\O{\Omega}
\def\d{\delta}
\def\p{\partial}

\def\half{\textstyle{\frac{1}{2}}}

\def\bdoc{\begin{document}}
\def\edoc{\end{document}}

\def\beq{\begin{equation}}
\def\eeq{\end{equation}}
\def\bea{\begin{eqnarray}}
\def\eea{\end{eqnarray}}
\def\ben{\begin{enumerate}}
\def\een{\end{enumerate}}
\def\la{\langle}
\def\ra{\rangle}
\def\a{\alpha}
\def\b{\beta}
\def\g{\gamma}\def\G{\Gamma}
\def\d{\delta}\def\D{\Delta}
\def\e{\epsilon}

\def\th{\theta}
\def\k{\kappa}
\def\l{\lambda}
\def\m{\mu}
\def\n{\nu}
\def\o{\omega}
\def\p{\pi}
\def\r{\rho}
\def\s{\sigma}
\def\t{\tau}
\def\L{{\cal L}}
\def\S{\Sigma }
\def\gsim{\; \raisebox{-.8ex}{$\stackrel{\textstyle >}{\sim}$}\;}
\def\lsim{\; \raisebox{-.8ex}{$\stackrel{\textstyle <}{\sim}$}\;}
\def\gtrsim{\gsim}
\def\lessim{\lsim}
\def\loc{{\rm local}}
\def\vm{v_{\rm max}}
\def\bh{\bar{h}}
\def\del{\partial}
\def\nab{\nabla}
\def\half{{\textstyle{\frac{1}{2}}}}
\def\fourth{{\textstyle{\frac{1}{4}}}}

\def\bD{{\bf D}}
\def\bE{{\bf E}}
\def\bF{{\bf F}}
\def\bB{{\bf B}}
\def\bP{{\bf P}}
\def\bV{{\bf v}}
\def\bv{{\bf v}}
\def\bx{{\bf x}}
\def\by{{\bf y}}
\def\bz{{\bf z}}
\def\ba{{\bf a}}
\def\bd{{\bf d}}
\def\bs{{\bf s}}
\def\bn{{\bf n}}
\def\bp{{\bf p}}

\def\O{\Omega}

\def\br{{\bf r}}
\def\bnab{{\bf \nab}}

\def\tE{\tilde{E}}
\def\tL{\tilde{L}}


\begin{document}

\title{Destroying a near-extremal Kerr black hole with a charged particle:
\\Can a test magnetic field serve as a cosmic censor?}

\author{Sanjar Shaymatov$^{1,2}$}
\email{sanjar@astrin.uz}
\author {Mandar Patil$^{3}$}
\email{mandar@iucaa.ernet.in}
\author{Bobomurat Ahmedov$^{1,2}$}
\email{ahmedov@astrin.uz}
\author{Pankaj S. Joshi$^{4}$}
\email{psj@tifr.res.in}

\affiliation{$^{1}$Institute of Nuclear Physics, Ulughbek,
Tashkent
              100214, Uzbekistan}
    \affiliation{$^{2}$Ulugh Beg Astronomical Institute,    Astronomicheskaya 33,
    Tashkent 100052, Uzbekistan}
\affiliation{$^{3}$Inter University Centre for Astronomy \&
Astrophysics, Post Bag 4, Pune 411007, India}
\affiliation{$^{4}$Tata Institute of Fundamental Research Homi
Bhabha Road, Mumbai 400005, India}

\date{\today}
\begin{abstract}
We investigate effect of a test magnetic field on the process
of destroying near-extremal Kerr black hole with a charged test
particle. It has been shown that it would be possible to throw
a charged test particle into the near extremal rotating black hole
and make it go past the extremality i.e. turn Kerr black hole into
the Kerr-Newmann naked singularity. Typically in an astrophysical
scenario black holes are believed to be surrounded by a
magnetic field.  Magnetic field although small, affects motion of
charged particles drastically due to the large Lorentz force,
as the electromagnetic force
is much stronger that the gravity. Thus a test magnetic field can affect
the process of destroying black holes and restore the cosmic censorship
in the astrophysical context. We show that a test magnetic field would act
as a cosmic censor beyond a certain threshold value. We try to gauge the magnitude
of the magnetic field by comparing its energy density with that
of the change in the curvature induced by the test particle.
We find that the magnetic field required in only as strong as or slightly stronger as
compared to the value for which its effect of the background geometry is
comparable to the tiny backreaction as that of the test particle. In such a
case however one has to take take into account effect of the magnetic
field on the background geometry, which is difficult to implement in the
absence of any exact near-extremal rotating magnetized black hole solutions.
We argue that magnetic field would still act as a cosmic censor.

\end{abstract}
 \maketitle

\section{Introduction}
\label{introduction}

In this paper we study a Gedanken experiment to destroy
a black hole with the infalling test particle. The infalling particle
would add to the mass, angular momentum and charge of the black hole and can make it go pass the
extremality, thus turning black hole into the naked singularity. Such a process was examined
by Wald for the first time, who found that it is impossible to overspin an extremal Kerr
black hole by throwing in a neutral test particle \cite{Wald}. If the angular momentum of the
infalling particle is large enough for the purpose of overspining the black hole,
it turns back before it could enter a black hole. Whereas if the particle enters the black hole
it would not add a sufficient amount of angular momentum to overspin the black hole.

Recently it was shown by Jacobsan and Sotiriou (JS)
\cite{Jacobson:2010iu} that it would be possible to destroy a
black hole with infalling test particle if we start with a
near-extremal configuration, rather than an extremal black hole as
in the case of Wald`s analysis. There is a narrow range of the
energy and angular momentum of the infalling particles for which
it would be possible for it to enter a black hole and also
overspin it past the extremality. It was also shown that it would
be possible to destroy near-extremal Reissner-Nordstr\"{o}m black
holes with charged test particle \cite{Hubney}. The process of
destroying rotating black hole with the charged test particle was
investigated in \cite{Alberto Saa}. In all these calculations it
was assumed that the test particle follows a geodesic motion and
effects of the conservative and dissipative backreaction were
ignored. There are investigations which suggest that the radiation
reaction and self-force would act as a cosmic censor
\cite{Enrico,Poisson,Rocha}. It was also shown that it would be
possible to destroy a regular black hole with the test particles
even when the backreaction effects are taken into account
\cite{Bambi}.

We approach this issue from a different perspective. Typically in
the astrophysical scenarios black holes are surrounded with the
magnetic field which would affect the motion of the charged
particle and thus influence the process of destroying black hole.
With this motivation in mind in this paper we investigate whether
or not a magnetic field could possibly serve as a cosmic censor.
For this purpose we introduce a test magnetic field on the Kerr
spacetime following a procedure that is described in \cite{Wald1}.
The Wald solution is recently extended to a black hole that is also 
moving at constant velocity in \cite{mbh1,mbh2,mbh3,mbh4}.
The magnetic field respects the axial symmetry of the Kerr
spacetime and it takes a constant value asymptotically at
infinity. We show that for large enough value of the magnetic field
it serves as a cosmic censor. We try to gauge the magnitude of the 
magnetic field required by comparing its backreaction on the 
background geometry as compared to that of a test particle.
The trace of the energy-momentum tensor of the magnetic
field is a good measure of its backreaction on the background Kerr
spacetime. We compute square root of the difference in the
Kretschmann scalar at the horizon between extremal and
near-extremal geometry, which is a fair indicator of the
backreaction of the test particle on the background spacetime.
We show that the backreaction of the magnetic field is as much 
as or slightly larger than that of the tiny backreaction of the 
test particle when it starts acting as the cosmic censor. 
Thus an extremely weak magnetic field is sufficient to restore the
cosmic censorship in the process of destroying near extremal Kerr
black hole. 

This analysis also suggests that we must in principle take into 
account the backreaction of the magnetic field on the background geometry
since change in the metric due to the magnetic field will be comparable to that of the test 
particle. It is difficult to implement it in the absence of any exact solution representing 
magnetized near extremal geometry. However we argue that our conclusions won`t change 
even after taking into account the backreaction of the magnetic field.

If black holes could indeed turn into the naked singularities i.e. if cosmic censorship
hypothesis \cite{Penrose} could be violated, it would have serious implications from
theoretical as well as observational perspective. Kerr and Kerr-Newmann naked singular
geometries are associated with the absence of global hyperbolicity and also with the
existence of the closed timelike curves. However it was suggested that string theory
could potentially resolve the naked singularities and all these issues would disappear
\cite{Horava} rendering their existence legal. From observational point of view it was shown that the near extremal
naked singularities can host ultra-high energy particle collisions and thus can serve as an
astrophysical probe of high energy physics \cite{Patil1,Patil2,Patil3,Patil4,Patil5}. We also
note that there are many investigations where it is demonstrated that the naked singularities can
form as an end-state of the continual gravitational collapse \cite{Joshi,Joshi1,Harada,Joshi2}.

We note that there are many papers analyzing motion of the charged particles in the magnetic
field around central objects, analytically as well as numerically. \citep[see,
e.g.][]{Prasanna,Dadhich,Frolov1,Aliev,Abdujabbarov,Abdujabbarov1,
Abdujabbarov2,Abdujabbarov3,Shaymatov}. The strength of the
magnetic field $B$ is estimated to be the $B_1 \sim 10^8 G$ for
stellar mass black holes of mass $ M \sim 10M_\odot$, and $B_2\sim
10^4 G $ for the supermassive black holes of mass $M \sim 10^9
M_\odot$ (\cite{Piotrovich}). The strength of the magnetic
fields near the event horizon of the black hole was also measured recently (\cite{Eatough,Shannon}).

In Sec.~\ref{Sec:charged-par} we describe the process of destroying
near-extremal Kerr black hole with a charged test particle.
In Sec.~\ref{Sec:mag-field} we introduce a test magnetic field on the
background of the Kerr black hole and analyze its effect
on the motion of the charged particle. In Sec.~\ref{Sec:test} we compare backreaction
of the test magnetic field with that of the test particle and analyze whether magnetic field
could possibly serve as a cosmic censor. We summarize our concluding remarks of the
obtained results in the Sec.~\ref{Sec:Conclusion}.

In this work we use a system of units in which $G=c=1$.

\section{Particle motion around Near-extremal Kerr black hole }\label{Sec:charged-par}

In this section we describe the process of destroying
near-extremal Kerr black hole with the charged test particle. We
calculate the range of the energy and angular momentum of the
particle for it to turn Kerr black hole into the Kerr-Newmann
naked singularity. We then analyze allowed range of the parameters
for which particle can enter the black hole.

We restrict our attention to the particles that follow geodesic
motion on the equatorial plane of the Kerr black hole with mass
$M$ and angular momentum $J$. There are two constants of motion
associated with the particle, namely conserved energy $\delta E$
and conserved angular momentum $\delta J$. We assume that these
quantities are much smaller compared to that of the black hole
$\delta E\ll M $, $\delta J\ll J $, so that the test particle
approximation holds good. Let $e$ be the charge associated with
the test particle which is also assumed to be small. We neglect
the radiation reaction and self-force. When particle enters the
black hole it adds to the mass, angular momentum and charge of the
black hole. The final mass, angular momentum and charge of the
black hole are given by $M+\d E$, $J+\d J$ and $e$ respectively.

If the particle were to turn a Kerr black hole into the
Kerr-Newmann naked singularity, following condition must hold:
\beq\label{naked} (M+\d E)< \left(\frac{J+\d J}{M+\d E}\right)^2 +
e^2. \eeq
This yields the lower bound on the angular momentum of the
particle
\beq\label{Jmin} \d J > \d J_{min}=(M^2-J)+2M \d E+\d E^2
-\frac{e^2}{2} . \eeq
As it was pointed out in \cite{Jacobson:2010iu}, null energy
condition on the matter the test particle consists of, puts an
upper bound on the angular momentum of the particle
\beq\label{Jmax} \d J < \d J_{max}=\frac{2Mr_+}{a} \d E, \eeq
where the two parameters, $a$ and $M$, denote specific angular
momentum and the mass of a Kerr black hole, respectively. In case
of the extremal black hole it turns out that $J_{max}<J_{min}$ and
thus it is not possible to overspin it.

In this paper we deal with the near-extremal black hole with
dimensionless spin parameter close to unity. We take
$J/M^2=a/M=1-2\e^2$, with $\epsilon \ll 1$ being a small
dimensionless parameter. Hereafter we set $M=1$. Maximum and
minimum values of $\d J$ are given by
\bea\label{minmax}
\d J_{min} &=& 2\e^2+2\d E + \d E^2 -\frac{e^2}{2},\\
\d J_{max}&=& (2+4\e)\d E,\label{jmax} \eea
\\
and the allowed range of $\d E$ is
\beq\label{dEe}
\left(2-\sqrt{2}\sqrt{1+\left(\frac{e}{2\e}\right)^2}\right)\e <\d
E<\left(2+\sqrt{2}\sqrt{1+\left(\frac{e}{2\e}\right)^2}\right)\e.
\eeq
%
 %
%
%

%
%
\begin{figure*}
a)  \includegraphics[width=0.45\textwidth]{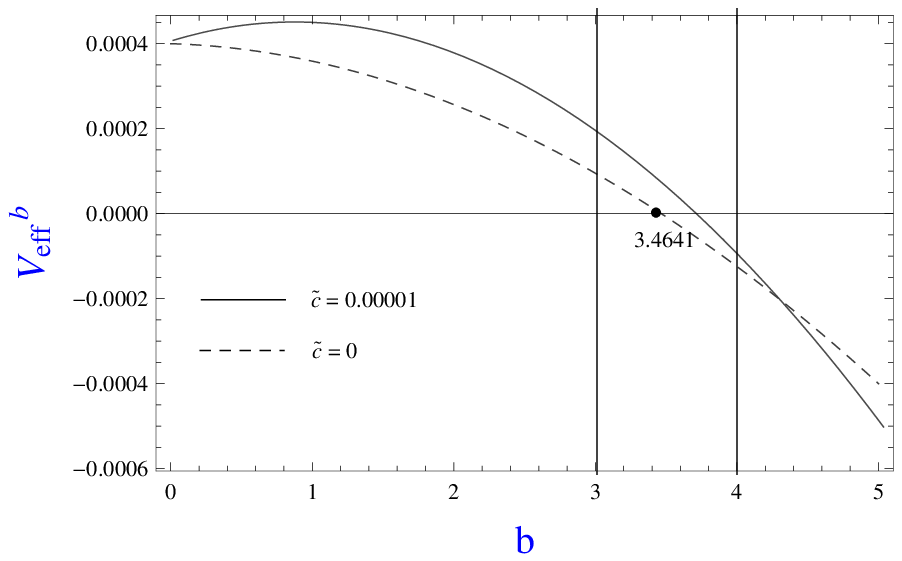}
b)  \includegraphics[width=0.45\textwidth]{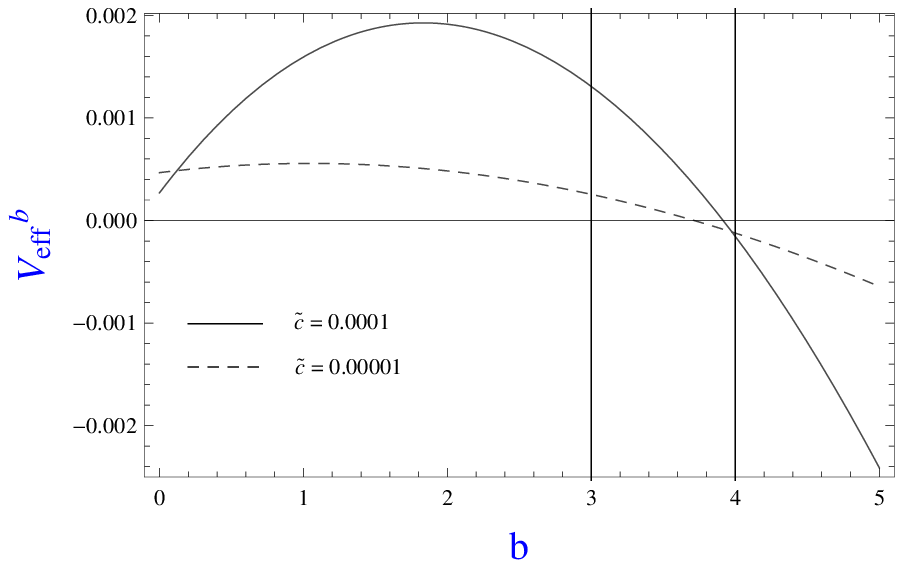} %

\caption{\label{Veffb} The dependence of value of the effective
potential at given maximum radius $r_{max}$ on the parameter $b$
is plotted here for the different values of charge parameter
$\tilde{c}$. The allowed range of $b$ increases as we increase the
charge $e$.}
\end{figure*}
The value of charge $e$ is taken to be small as compared to
$\epsilon$. Thus we have $\d E$ of the order of $\e$. For the
given $\d E$ we get $\d J\sim \d E$. Thus we have $\d E\ll M$ and
$\d J\ll J$ and the particle under consideration can be thought of
as a test particle.

We must ensure that the particle in the allowed range of the energy and angular
momentum starting from a distant location indeed enters the black hole if it were to
turn it into a naked singularity. Thus we need to understand the geodesic motion
of the particle. As stated earlier we focus on the particle that is restricted
to move on the equatorial plane.
%
%
The motion of the particle in the radial direction can be
described in terms of the effective potential as stated in the
equation below. Thus we must analyze the behavior of the effective
potential to understand whether or not particle enters the black
hole:
\beq\label{Eff1} \frac{\dot{r}^2}{2} + V_{\rm eff}(r,\tilde{\d
E},\tilde{\d J})=0, \eeq
where $\tilde{\d E}=\d E/m$ and $\tilde{\d J}=\d J/m$, and $m$ is
the rest mass of the particle.

For the chosen value of the energy $\d E$ one can write the
allowed range of the angular momentum of the body falling
into the black hole as
\beq\label{rangeJ} (2+3\e)\tilde{\d E}-\frac{e^2}{2m}<\tilde{\d
J}<(2+4\e)\tilde{\d E}. \eeq

As we have already mentioned, the initial black hole is nearly
extremal,
%
but now we can be
somewhat more quantitative. For example, we take $\e=10^{-2}$. The
spin parameter of the Kerr black hole in this case is given by
$a=0.9998$. We can imagine even smaller values of $\e$ in
principle.

We parametrize the range of allowed specific angular momentum in the following way
\beq\label{rangeJ} \tilde{\d J}=(2+b~\e)~\tilde{\d
E}-(4-b)~\tilde{c}, \eeq
where $b\in[3,4]$ and $\tilde{c}=e^2/2m$ is charge parameter.

We now investigate the effective potential for the radial motion
of the particle. The charged particle is assumed to start from a
distant location falling in towards the black hole. We must make
sure that $V_{eff}<0$ everywhere outside the horizon so that the
particle enters the black hole. The effective potential that
appears in (\ref{Eff1}), is given by
\begin{widetext}
\begin{align}
V_{eff} =&-\frac{\tilde{\d
E}^2}{2}\bigg[1-\frac{3+b\tilde{c}(b-4)+4b\e+(4+b^2)\e^2}{r^2}+\frac{2-2b\tilde{c}(b-4)+4b\e-4\tilde{c}(4-b)\e+2(4+b^2)\e^2}{r^3}
 \cr  &-\gamma\left(\frac{8b+(4-b)^2\gamma -32}{4r^2}-\frac{4b+(4-b)^2\gamma -16}{2r^3}\right)\bigg]
, \label{Eff2}
\end{align}
\end{widetext}
where $\gamma=\tilde{c}/\e$. Here we assume that the specific
energy of the infalling particle is large $\d E>>1$ as in
\cite{Jacobson:2010iu}.

We compute the maximum value of the effective potential attained
at a location outside the horizon and write it as
$V_{eff}=-\tilde{\d E}^2 V_{eff}^{b}/2$. For particle to enter the
event horizon  $V_{eff}^{b}$ must be positive at the radial
location stated above where effective potential attains maximum.
For given value of $\tilde{c}$ the parameter $b$ must take a value
in the range $(3,b_{cr})$ as it can be seen in Fig 1. The upper
critical values of the parameter $b_{cr}$ for the different values
of charge parameter $\tilde{c}$ are tabulated in Table 1.
\begin{table}[ht]
\caption{\label{table1} The values of parametrization parameter
$b_{cr}$ for the different values of charge parameter $\tilde{c}$
for the given value of non-extremality $\e$.} \centering
\begin{tabular}{c c c c c c c}
\hline\hline $\e$ &~ $0.01$ &~ $0.01$ &~ $0.01$ &~ $0.01$ &~
$0.01$ &~ $0.01$ \\ [0.5ex]
$\tilde{c}$&~ $10^{-3}$ &~ $10^{-4}$ &~ $5\times 10^{-5}$ &~ $10^{-5}$ &~ $5\times10^{-6}$ &~ 0 \\
$b_{cr}$ &~ 3.9863 &~3.8846  &~  3.8034 &~ 3.6093 &~ 3.5558 &~
3.4641
\\ [1ex] \hline\hline
\end{tabular}
\end{table}

The range of parameter $b$ is given in $[3,4]$. As shown in
\cite{Jacobson:2010iu} the allowed range for the uncharged
particle to enter the black hole is given by $[3,\sqrt{12}]$. As
it can be seen in Fig \ref{Veffb}, the effective potential
$V_{eff}^{b}$ at $r_{max}$ is positive in this range. Here we
analyze the process of destroying Kerr black hole with charged
particle. The allowed range of parameter $b$ for which test
particle enters the near-extremal black hole and turns it into the
Kerr-Newmann naked singularity increases as we increase the charge
parameter. As it can be seen from Fig \ref{Veffb} the intersection
point of effective potential curve moves towards $b=4$.
Consequently it becomes easier to destroy the black hole if
incoming particle is charged.

In this section we described the process of destroying
near-extremal Kerr black hole with a charged test particle. In the
next section we introduce a test magnetic field around the Kerr
black hole and analyze its effect on the process of destroying
black hole.
\section{Particle motion around black hole in a magnetic field}\label{Sec:mag-field}

We analyze the effect of magnetic field in the process of
destroying Kerr black hole with a charged particle. We describe a
process to set up a test magnetic field on the spacetime
containing Kerr black hole. The magnetic field takes a constant
value at infinity and is oriented along the axis of symmetry of
the Kerr geometry. The motion of the charged particle could be get
influenced by the test magnetic field and it can also affect the
process of destroying black hole. We try to understand whether or
not magnetic field can stop particles with the appropriate values
of geodesic parameters from entering the black hole and thus it
would serve as a cosmic censor.

The metric of the Kerr geometry in the Boyer-Lindquist coordinates
is given by
\begin{align}
ds^2 =&-\left(\frac{\Delta-a^2\sin^2\theta}{\Sigma}\right)dt^2
-\frac{2a\sin^2\theta(r^2+a^2-\Delta)}{\Sigma}dtd\phi \cr
&+\frac{(r^2+a^2)^2-\Delta a^2\sin^2\theta}{\Sigma}\sin^2\theta
d\phi^2 +\frac{\Sigma}{\Delta}dr^2+\Sigma d\theta^2,
\label{eq:metric}
\end{align}
where $\Sigma=r^2+a^2\cos^2\theta$ and $\Delta=r^2+a^2-2Mr$.
%

Kerr space-time admits two Killing vectors,
$\xi^{\alpha}_{(t)}=(\partial/\partial t)^{\alpha}$ and
$\xi^{\alpha}_{(\varphi)}=(\partial/\partial \phi)^{\alpha}$. They
satisfy two Killing equations \cite{Wald1}
\begin{eqnarray}
\xi_{\alpha; \beta}+\xi_{\beta; \alpha}= 0 ,
\end{eqnarray}
which can be written in the form $ \Box \xi^\alpha=0$. The
vacuum Maxwell equations for vector potential $A_\alpha$ have the
same form $ \Box A^\alpha=0$ in the Lorentz gauge. Thus the vector
potential can be described as combination of the Killing vectors
\cite{Wald1}
\begin{eqnarray}\label{4-pot}
A^{\alpha}=C_1 \xi^{\alpha}_{(t)}+ C_2 \xi^{\alpha}_{(\varphi)} .
\end{eqnarray}

We select the integration constants $C_1=0$ and $C_2=B/2$ and the
vector potential can be written as
\begin{equation}
 A^{\alpha}= B/2 \xi^{\alpha}_{(\varphi)}
\end{equation}
It can be shown from the asymptotic properties that $B$ turns out to be the
magnetic field at infinity which is uniform and oriented along the axis of symmetry.

The covariant
components of the 4-vector potential of the electromagnetic field
will take the form
\begin{eqnarray}
\label{pot.1} A_t&=&-\frac{1}{2\Sigma}\bigg\{ a
B\bigg[\Delta\big(1+cos^2\theta\big)+
\big(r^2-a^2\big)\sin^2\theta\bigg] \nonumber\\
&& - 2aB\big(\Sigma-2Mr\big)\bigg\},\nonumber\\
\nonumber\\A_{r}&=& A_{\theta}=0\ ,\nonumber\\
\nonumber\\
A_{\varphi}&=&\frac{1}{\Sigma}\bigg\{\frac{B}{2}\bigg[\Delta
a^2\big(1+cos^2\theta\big)+
 r^4-a^4\bigg] \nonumber\\
 && - 2 Q MB a^3\bigg\}\sin^2\theta\ .
\end{eqnarray}
%

%

We write down the conserved quantities for particle motion in the
equatorial plane
\begin{eqnarray}
\label{En1} \pi_{t}&=& -g_{\mu\nu}(\xi_{t})^{\mu}\pi^{\nu}=
g_{tt}\pi^{t} + g_{t\varphi}\pi^{\varphi}+e A_{t},\\
\nonumber\\
 \label{Ln1}
\pi_{\varphi}&=&
-g_{\mu\nu}(\xi_{\varphi})^{\mu}\pi^{\nu}=g_{\varphi t}\pi^{t} +
g_{\varphi\varphi}\pi^{\varphi}+e A_{\varphi},
\end{eqnarray}
where $\pi^{\nu}$ is the four velocity defined by
$\pi^{\nu}=\frac{dx^{\nu}}{d\tau}$, $\tau$ is the proper time for
timelike geodesics.

Solving equations (\ref{En1}) and (\ref{Ln1}) we write down the
equation of motion for the charged particle motion in the Maxwell
field around black holes
\begin{eqnarray}
\label{tt} \pi^{t}&=&\frac{1}{r^2}\left[ a \big(\pi_{\varphi}+a\pi_{t}\big)+\frac{r^2+a^2}{\Delta}P\right], \\ \nonumber \\
\label{ff}
\pi^{\varphi}&=&\frac{1}{r^2}\left[ \big(\pi_{\varphi}+a\pi_{t}\big)+\frac{a}{\Delta}P\right] ,\\ \nonumber\\
\label{rr}
\left(\pi^{r}\right)^2&=&\frac{P^2-\Delta\big[r^2+\big(\pi_{\varphi}+a\pi_{t}\big)^2\big]}{r^4},
\end{eqnarray}
where $P=(r^2+a^2)(-\pi_{t})-a \pi_{\varphi}$.

The effective potential for the radial motion of charged particle
at the equatorial plane $\theta=\pi/2$ of the Kerr black hole
placed in an external magnetic field is given by
%
%
%
\begin{eqnarray}\label{eff}
V_{eff}(r)&=&\frac{\Delta\big[r^2+\big(\pi_{\varphi}+a\pi_{t}\big)^2\big]-P^2}{2r^4}
\ .
\end{eqnarray}
Now we discuss and analyze charged particle motion around a Kerr
black hole immersed in a uniform magnetic field. Again we analyze
the effective potential. Using the Eq.~(\ref{eff}), the effective
potential for radial motion can be given as
\begin{align}
\label{Veff1} V_\mathrm{eff}
&=-\frac{1}{2r^2}\left[\left(r^2+a^2+\frac{2Ma^2}{r}\right)\left(\tilde{\d
E}^2-\frac{\beta^2}{4M^2}\Delta\right)\right. \cr
&\left.-\left(1-\frac{2M}{r}\right)\tilde{\d
J}^2-\frac{4Ma~\tilde{\d E}~\tilde{\d J}}{r} -\Delta
\left(1-\frac{\beta ~\tilde{\d J}}{M}\right)\right],
\end{align}
where 
the magnetic parameter $\beta=eBM/m$ measures the influence of the
magnetic field on charged particle motion.
\begin{figure*}
\centering
%
a)  \includegraphics[width=0.45\textwidth]{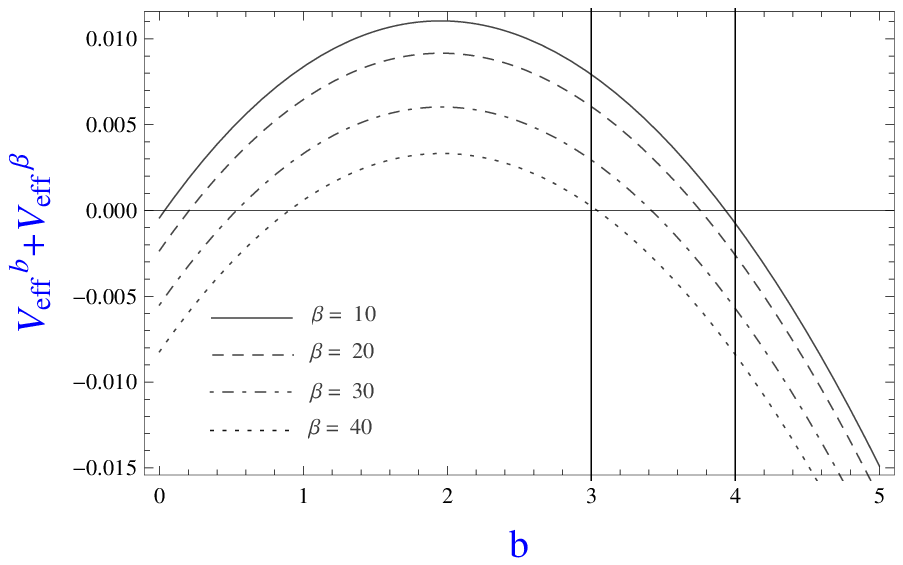}
b)  \includegraphics[width=0.45\textwidth]{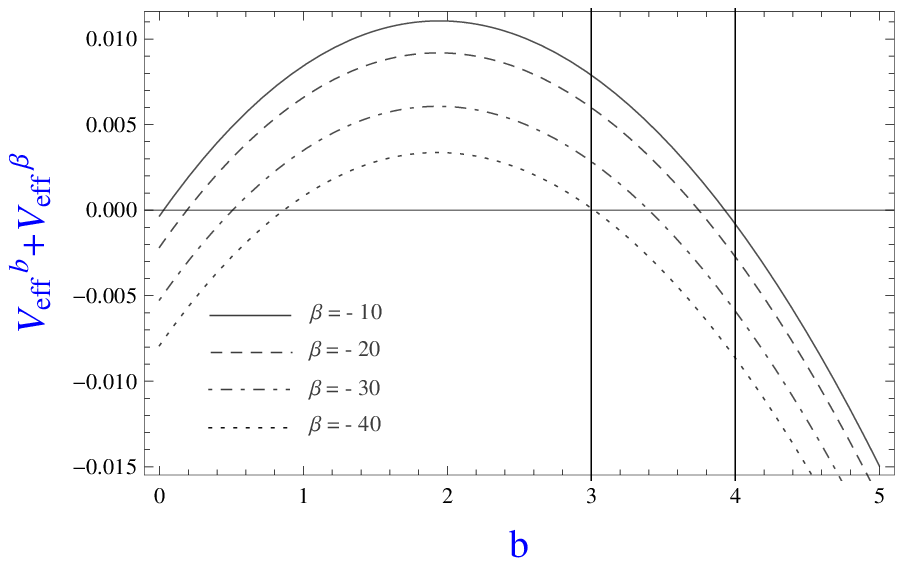}

%
\caption{\label{beta} The dependence of the effective potential at
the maximum point $r_{max}$ near extremal rotating black hole
placed in a magnetic field of strength $B$ on the parametrization
parameter $b$ for both negative $\beta<0$ and positive $\beta>0$
cases for the different values of magnetic parameter $\beta$ in
the case when charge parameter $\tilde{c}=10^{-3}$.}
\end{figure*}

We study the effective potential in order to understand the effect
of magnetic field on the process of destroying black hole. We
would like to be sure that the particle with the energy and
angular momentum in the appropriate range as described in the
earlier section will start from a infinity and fall towards the
black hole without encountering any turning point.

The effective potential can be written in the following form
%
\begin{align}
V_{eff} =&-\frac{\tilde{\d
E}^2}{2}\bigg(V_{eff}^{b}+V_{eff}^{\beta}\bigg), \label{Veff2}
\end{align}
%
where $V_{eff}^{\beta}$ is defined as follows
\begin{widetext}
\begin{align}
V_{eff}^{\beta} =&~\beta\bigg[\frac{2+b \e}{\tilde{\d E}}-\frac{8
\tilde{c}-2b \tilde{c}+\beta(1-4 \e^2)}{2\tilde{\d E}^2}-\frac{(4
+2b \e)/\tilde{\d E}-(8 \tilde{c} -2b \tilde{c})/\tilde{\d
E}^2}{r}+\frac{(16b\tilde{c}-64 \tilde{c} + 3\beta)/\tilde{\d
E}^2}{4r^2}\cr +&\frac{(2 +b\e-8 \e^2)/\tilde{\d E}-(4b \tilde{c}-
16 \tilde{c}+2\beta)(\e/\tilde{\d E})^2 }{r^2}
   -\frac{\beta(1-8\e^2)/\tilde{\d
E}^2}{2r^3}\bigg] . \label{Veff3}
\end{align}
\end{widetext}
We have expanded the potential (\ref{Veff1}) out to second order
in $\e$.
%
%
%


 We now try to understand the effect of the test
magnetic field on the motion of the test particle and see whether or not it can
serve as a cosmic censor in a process of destroying black hole.

For a particle with a given mass and charge, the magnetic
parameter $\beta$ increases with the increasing magnetic field.
For low values of the magnetic fields and parameter $\beta$ it
will not be possible for magnetic field to prevent particles from
entering the black hole and turning it into the naked singularity.
However with increase of the magnetic field and parameter $\beta$
the motion of the charged particles is significantly affected. We
plot effective potential at the maximum
$V_{eff}^{b}+V_{eff}^{\beta}$ as a function of parameter $b$. The
allowed range of the angular momenta for which it is possible to
destroy black hole is given by $(3,b_{cr})$ where
$V_{eff}^{b}+V_{eff}^{\beta}$ is positive. As we can see from the
Fig 2, that $b_{cr}$ tends to decrease as we increase the
magnitude of parameter $\beta$. At a certain critical value of
$\beta$ we have $b_{cr}=3$. Beyond this value it is not possible
for the charged particle to enter the black hole, and the test
magnetic field serves as a cosmic censor.

The effective potential for the radial potential is plotted in Fig
3. When the magnetic field is zero, the maximum of effective
potential is negative, thus allowing infalling particle to enter
the black hole. As we increase the magnetic field the height of
maximum tends to increase. Beyond certain value of the magnetic
field maximum value crosses zero and is positive. Thus the
infalling particle will turn back and will not be able to enter
the black hole.

We have shown that if the magnetic field is sufficiently large it
can prevent infalling particle from entering the black hole and
thus could in principle serve as a cosmic censor. In the next section we 
try to gauge how large is the critical magnetic field by comparing its backreaction 
with that of the test particle.

\begin{figure*}
\centering
a)  \includegraphics[width=0.3\textwidth]{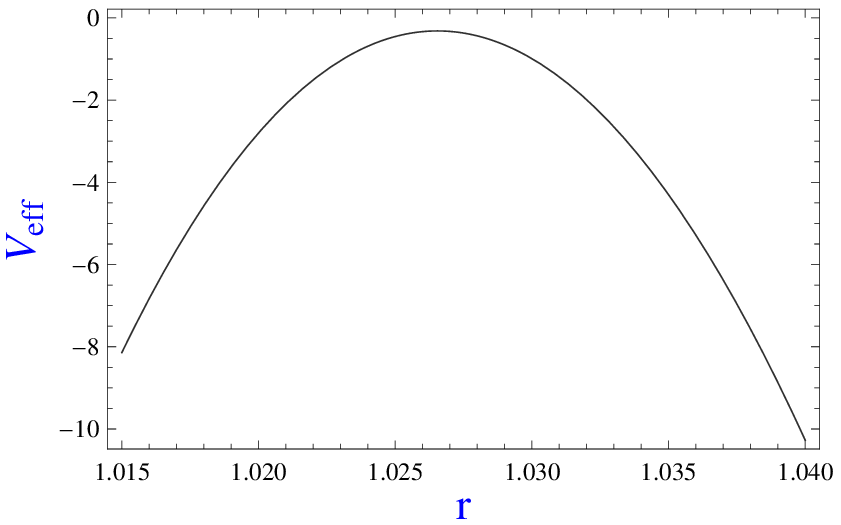}
b)  \includegraphics[width=0.3\textwidth]{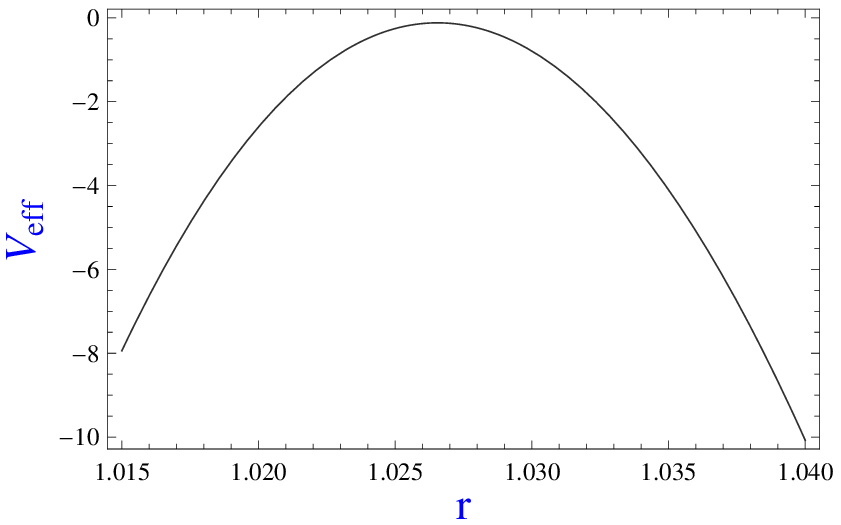}
c)  \includegraphics[width=0.3\textwidth]{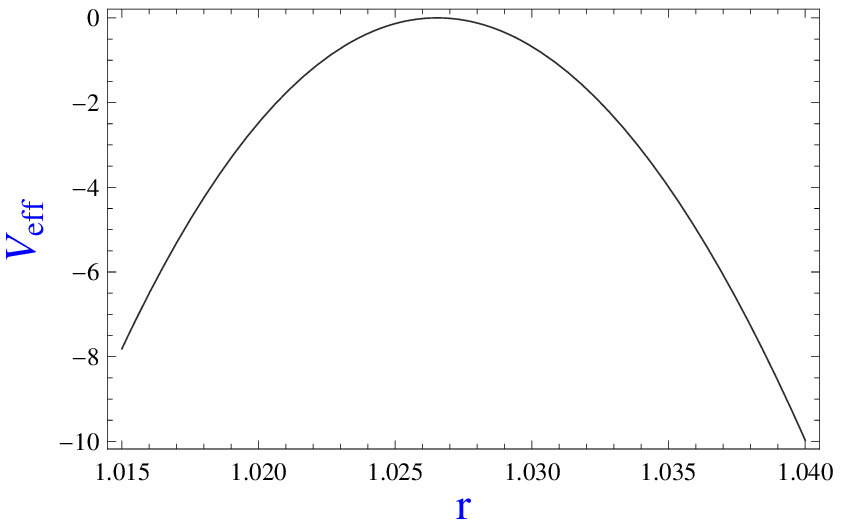}

d)  \includegraphics[width=0.3\textwidth]{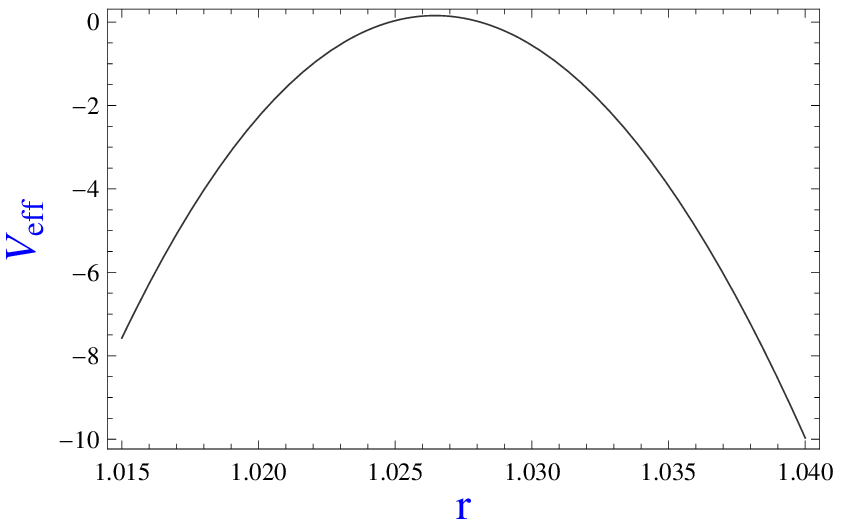}
e)  \includegraphics[width=0.3\textwidth]{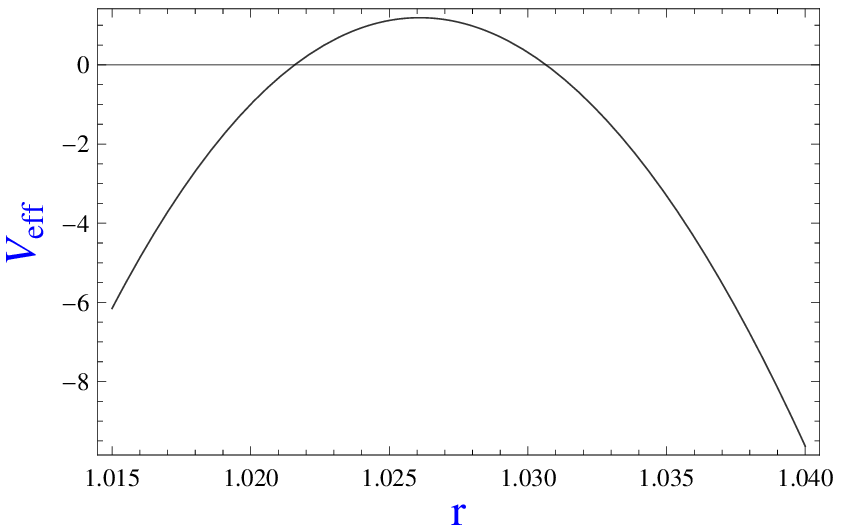}
f)  \includegraphics[width=0.3\textwidth]{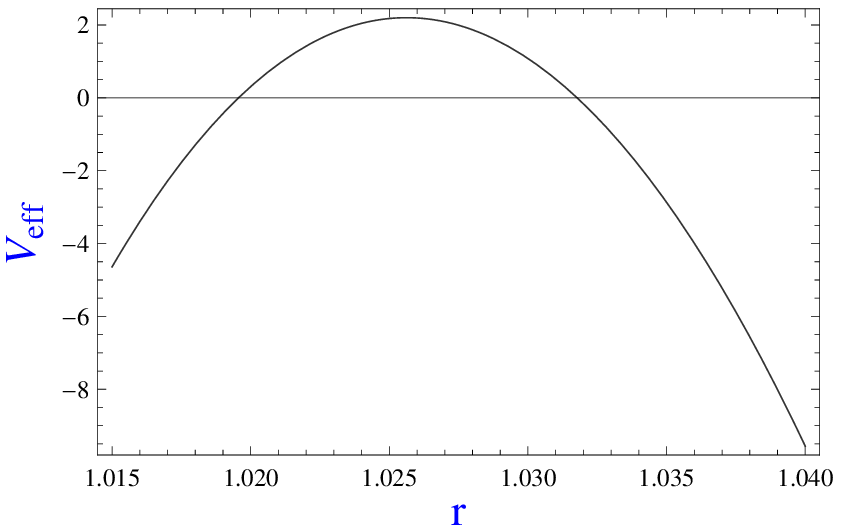}
\caption{\label{Veffr} Radial dependence of the effective
potential on the radial motion of the charged particle moving
around the near extremal rotating black hole immersed in a
magnetic field of strength $B$ for the different values of
magnetic parameter $\beta$. For this figure, $\beta=0$ (a),
$\beta=0.01$ (b), $\beta=0.1$ (c), $\beta=1$ (d), $\beta=5$ (e),
and $\beta=10$ (f) in the case when charge parameter
$\tilde{c}=10^{-3}$.}
\end{figure*}

\section{Analyzing the bacreaction of magnetic field on background spacetime}\label{Sec:test}

We have shown that the test magnetic field can potentially serve
as the cosmic censor preventing a particle that can turn
near-extremal black hole into a naked singularity in the absence
of magnetic field, from entering the black hole. In this section we try
to understand how large is the threshold magnetic field. As we describe below we do that by
comparing the strength of the perturbation of the magnetic field 
on the background spacetime with that of the test particle. 

The effect of the test particle on the background spacetime can be understood
in terms of the change in the Kretschmann scalar between extremal and near-extremal geometry
at the horizon. The backreaction of the magnetic field can be expressed in terms of the
trace of the energy momentum tensor of the test magnetic field calculated at the horizon of the
black hole. Trace of the energy momentum tensor has units of density, whereas Kretschmann scalar
has units of the square of density. Thus to compare the backreaction of test particle on the background geometry
to that of test magnetic field we compare square root of change in the Kretschmann scalar to the trace of the
energy momentum tensor.

Kretschmann scalar for the Kerr metric is given by the following
expression
\begin{eqnarray}\label{K}
K&=&R^{\alpha\beta\mu\nu}R_{\alpha\beta\mu\nu}=\frac{48M^2\left(r^2-a^2x^2\right)}{\big(r^2+a^2x^2\big)^6}\nonumber\\
&&\times \bigg((r^2+a^2x^2)^2-16a^2r^2x^2\bigg) \, ,
\end{eqnarray}
where $x= \cos\theta$.

We calculate the square root of the difference of Kretschmann
scalar at the horizon for extremal and near-extremal geometries,
subtract and take a square root:
%
%
\begin{align}\label{K1}
K&=(K_1-K_2)^{1/2}\cr & =6\sqrt[4]{2} \bigg[
2\sqrt{2}\bigg(\frac{1-7x^6+35x^4-21x^2}{(1+x^2)^7}\bigg)^{1/2}
(\epsilon)^{1/2}\cr &-
\frac{7-2x^{10}+47x^8-224x^6+434x^4-182x^2}{(1+x^2)^8\bigg(\frac{1-7x^6+35x^4-21x^2}{(1+x^2)^7}\bigg)^{1/2}}(\epsilon)^{3/2}\bigg]
\ , \cr
\end{align}
which measures the change in the curvature in the background
spacetime due to the test particle.

We now calculate the trace of the energy-momentum tensor of the
test magnetic field. The energy-momentum tensor of the
electromagnetic field is given by
\begin{eqnarray}
T^{\mu\nu}=\frac{1}{4
\pi}(F^{\mu}_{\sigma}F^{\nu\sigma}-\frac{1}{4}g^{\mu\nu}F_{\sigma\lambda}F^{\sigma\lambda}).
\label{eq:tensor}
\end{eqnarray}
The trace of the energy-momentum tensor of the magnetic field is given by
\begin{widetext}
\begin{align}\label{trace}
T&=\frac{B^2}{\Sigma^6 (2 M r+\Sigma)}\left[\frac{ (2 M r+\Sigma)
\left(\Xi-4 M^2 r^2+\Delta (2 M r+\Sigma)\right)}{2 \Xi \Sigma^2}
\left\{-\frac{4 \Sigma^4 x^2 \left(\left(\Xi-4 a^2 M r\right)
\left(a^2+r^2\right)-a^2 \Delta \Sigma
\left(1-x^2\right)\right)^2}{\Upsilon^2}\right.\right.\cr
&\left.\left.-a^2 M^2 \left(\frac{\Xi \Sigma \left(1+x^2\right)
\left(r^2-a^2 x^2\right)}{2 M r+\Sigma}+r \left(1-x^2\right)
\left(\Xi r-4 a^2 M r^2-2 r^3 \Sigma+a^2 (M-r) \Sigma
\left(1+x^2\right)\right)\right)^2\right\}\right.\cr
&\left.+2 \bigg\{\left(\Xi-4 M^2 r^2+2 \Sigma^2-\Delta (2 M
r+\Sigma)\right) x \left(a^6 M^2 r^2 \left(x^2-1\right)
\left(1+x^2\right)^2\right.\right.\cr
&+\left.\left.\left.\frac{\Sigma^2 \left(\Xi r-4 a^2 M r^2-2 r^3
\Sigma+a^2 (M-r) \Sigma
\left(1+x^2\right)\right)^2}{\Upsilon^2}\right)-\frac{1}{\Xi}\left(\Xi+4
M^2 r^2-2 \Sigma^2-\Delta (2 M r+\Sigma)\right)
\left(1-x^2\right)\right.\right.\cr
&\times\left.\left. \left(M^2 r^2 \left(\Xi r-4 a^2 M r^2-2 r^3
\Sigma+a^2 (M-r) \Sigma
\left(1+x^2\right)\right)^2\right.\right.\right.\cr
&\left.\left.\left.-\frac{a^2 x \big(\Xi \left(r^2 \Sigma+a^2 (4 M
r+\Sigma)-1\right) -a^2 (2 M r+\Sigma) \left(4 a^2 M r+4 M
r^3+\Delta \Sigma(1-
x^2)\right)\big)^2}{\Upsilon^2}\right)\right\}\right].
\end{align}
\end{widetext}
Here we have $\Upsilon=\left[a^2+2 r^2+a^2(2x^2-1)\right]$. At the
horizon trace is given by
\begin{widetext}
\begin{align}\label{tracer}
T=g_{\mu\nu}T^{\mu\nu}=\frac{4 B^2 \left( \left(1+x^2\right)^4
\left(3-2 x-5 x^2-3 x^3+x^4+x^6+x^7\right)+x \left(x^2-1\right)^2
\left(3+x^2\right)\right)}{\left(1+x^2\right)^8
\left(3+x^2\right)},
\end{align}
\end{widetext}
%

%
\begin{table*}[ht]
\caption{\label{table2} $b_{cr}$ is tabulated for the different
values of $\beta>0$. The ratio of magnetic field to the critical
value is also specified. Magnetic field starts acting as a cosmic
censor i.e. $b_{cr}<3$ when the magnetic field is around $10$ times larger
than the critical value.} \centering
\begin{tabular}{c c c c c c c c}
\hline\hline $$ &~~ $$ &~~ $$ &~~ $B=0$ &~~ $$ &~~
$$ &~~$$
\\
[0.5ex]\hline

$\e$ &~~ $0.01$ &~~ $0.01$ &~~ $0.01$ &~~ $0.01$ &~~ $0.01$ &~~ $0.01$ \\
[0.5ex]
$\tilde{c}$&~ $10^{-3}$ &~~ $10^{-4}$ &~~ $5\times 10^{-5}$ &~~ $10^{-5}$ &~~ $5\times10^{-6}$ &~~ $10^{-6}$  \\
$b_{cr}$ &~~ 3.986380697 &~~ 3.884691117  &~~ 3.803417951 &~~
3.609301958 &~~ 3.555779171 &~~ 3.5141640012 \\
[0.5ex]\hline$$ &~~ $$ &~~ $$ &~~ $B=10^{1} B_{cr}$ &~~ $$ &~~
$$ &~~
$$ \\ [0.5ex]\hline

$\beta$ &~~ $3.84 \times10^{1}$ &~~ $1.21\times10^{1} $  &~~
$8.59$
&~~ $3.84$ &~~ $2.72 $ &~~ $1.21 $\\
$b_{cr}$ &~~ 2.982423607 &~~ 3.016905974  &~~ 3.085491813 &~~
3.358287139 &~~ 3.359977248 &~~ 3.362780518 \\ [0.5ex]\hline
$$
&~~ $$ &~~ $$ &~~ $B= B_{cr}$ &~~ $$ &~~
$$ &~~
$$ \\ [0.5ex]\hline

$\beta$ &~~ $3.84$ &~~ $1.21 $  &~~ $8.59\times10^{-1}$
&~~ $3.84\times10^{-1}$ &~~ $2.72 \times10^{-1}$ &~~ $1.21 \times10^{-1}$\\
$b_{cr}$ &~~ 3.979232348 &~~ 3.880141686  &~~ 3.800030724 &~~
3.608611326 &~~ 3.554863065 &~~ 3.513952567 \\ [0.5ex]\hline
$$
&~~ $$ &~~ $$ &~~ $B=5\times10^{-1} B_{cr}$ &~~ $$ &~~
$$ &~~
$$ \\ [0.5ex]\hline

$\beta$ &~~ $19.2\times10^{-1}$ &~~ $6.05 \times10^{-1}$  &~~
$42.95\times10^{-2}$
&~~ $19.2\times10^{-2}$ &~~ $13.6 \times10^{-2}$ &~~ $6.05 \times10^{-2}$\\
$b_{cr}$ &~~ \textcolor{black}{3.984852247} &~~
\textcolor{black}{3.884176799} &~~
\textcolor{black}{3.803260456} &~~ 3.609582496 &~~ 3.556037015 &~~ \textcolor{black}{3.514451688} \\
[0.5ex]\hline
$$
&~~ $$ &~~ $$ &~~ $B=3\times10^{-1} B_{cr}$ &~~ $$ &~~
$$ &~~
$$ \\ [0.5ex]\hline
$\beta$ &~~ $11.52\times10^{-1}$ &~~ $3.63 \times10^{-1}$  &~~
$25.77\times10^{-2}$
&~~ $11.52\times10^{-2}$ &~~ $8.16 \times10^{-2}$ &~~ $3.63 \times10^{-2}$\\
$b_{cr}$ &~~ \textcolor{black}{3.985953499} &~~
\textcolor{black}{3.884803702} &~~
\textcolor{black}{3.803690923} &~~ 3.609620127 &~~ 3.556105241 &~~ 3.514430845 \\
[0.5ex]\hline
$$
&~~ $$ &~~ $$ &~~ $B=2\times10^{-1} B_{cr}$ &~~ $$ &~~
$$ &~~
$$ \\ [0.5ex]\hline
$\beta$ &~~ $7.68\times10^{-1}$ &~~ $2.42 \times10^{-1}$  &~~
$17.18\times10^{-2}$
&~~ $7.68\times10^{-2}$ &~~ $5.44 \times10^{-2}$ &~~ $2.42 \times10^{-2}$\\
$b_{cr}$ &~~ \textcolor{black}{3.986259107} &~~
3.884906431 &~~
3.803722289 &~~ 3.609563974 &~~ 3.556053615 &~~ 3.514373287 \\
[0.5ex]\hline
$$
&~~ $$ &~~ $$ &~~ $B=10^{-1} B_{cr}$ &~~ $$ &~~
$$ &~~
$$ \\ [0.5ex]\hline
$\beta$ &~~ $3.84\times10^{-1}$ &~~ $1.21 \times10^{-1}$  &~~
$8.59\times10^{-2}$
&~~ $3.84\times10^{-2}$ &~~ $2.72 \times10^{-2}$ &~~ $1.21 \times10^{-2}$\\
$b_{cr}$ &~~ 3.986401494 &~~ 3.884868886  &~~ 3.803631261 &~~
3.609457903 &~~ 3.555944907 &~~ 3.514284328 \\ [0.5ex]\hline
$$
&~~ $$ &~~ $$ &~~ $B=10^{-2} B_{cr}$ &~~ $$ &~~
$$ &~~
$$ \\ [0.5ex]\hline

$\beta$ &~~ $3.84\times10^{-2}$ &~~ $1.21 \times10^{-2}$  &~~
$8.59\times10^{-3}$
&~~ $3.84\times10^{-3}$ &~~ $2.72 \times10^{-3}$ &~~ $1.21 \times10^{-3}$\\
$b_{cr}$ &~~ 3.986390120 &~~ 3.884715198  &~~ 3.803444782 &~~
3.609319796 &~~ 3.555787247 &~~ 3.514177445 \\ [0.5ex]\hline
$$
&~~ $$ &~~ $$ &~~ $B=10^{-4} B_{cr}$ &~~ $$ &~~
$$ &~~
$$ \\ [0.5ex]\hline

$\beta$ &~~ $3.84\times10^{-4}$ &~~ $1.21 \times10^{-4}$  &~~
$8.59\times10^{-5}$
&~~ $3.84\times10^{-5}$ &~~ $2.72 \times10^{-5}$ &~~ $1.21 \times10^{-5}$\\
$b_{cr}$ &~~ 3.986380799 &~~ 3.884692523  &~~ 3.803418226 &~~
3.609302363 &~~ 3.555779365 &~~ 3.514164137 \\ [0.5ex]\hline
$$
&~~ $$ &~~ $$ &~~ $B=10^{-6} B_{cr}$ &~~ $$ &~~
$$ &~~
$$ \\ [0.5ex]\hline

$\beta$ &~~ $3.84\times10^{-6}$ &~~ $1.21 \times10^{-6}$  &~~
$8.59\times10^{-7}$
&~~ $3.84\times10^{-7}$ &~~ $2.72 \times10^{-7}$ &~~ $1.21 \times10^{-7}$\\
$b_{cr}$ &~~ 3.986380698 &~~ 3.884691119  &~~ 3.803417954 &~~
3.609301961 &~~ 3.555779173 &~~ 3.5141640010

\\ [1ex] \hline\hline
\end{tabular}
\end{table*}
\begin{table*}[ht]
\caption{\label{table3} $b_{cr}$ is tabulated for the different
values of $\beta<0$. The ratio of magnetic field to the critical
value is also specified. Magnetic field starts acting as a cosmic
censor i.e. $b_{cr}<3$ when the magnetic field is  around $10$
times larger than the critical value.} \centering
\begin{tabular}{c c c c c c c c}
\hline\hline $$ &~~ $$ &~~ $$ &~~ $B=0$ &~~ $$ &~~
$$ &~~$$
\\
[0.5ex]\hline

$\e$ &~~ $0.01$ &~~ $0.01$ &~~ $0.01$ &~~ $0.01$ &~~ $0.01$ &~~ $0.01$ \\
[0.5ex]
$\tilde{c}$&~ $10^{-3}$ &~~ $10^{-4}$ &~~ $5\times 10^{-5}$ &~~ $10^{-5}$ &~~ $5\times10^{-6}$ &~~ $10^{-6}$  \\
$b_{cr}$ &~~ 3.986380697 &~~ 3.884691117  &~~ 3.803417951 &~~
3.609301958 &~~ 3.555779171 &~~ 3.5141640012 \\
[0.5ex]\hline$$ &~~ $$ &~~ $$ &~~ $B=10^{1}B_{cr}$ &~~ $$ &~~
$$ &~~
$$ \\ [0.5ex]\hline

$\beta$ &~~ $3.84 \times10^{1}$ &~~ $1.21 \times10^{1}$  &~~
$8.59$
&~~ $3.84$ &~~ $2.72$ &~~ $1.21$\\
$b_{cr}$ &~~ 2.861504023 &~~ 3.008642024  &~~ 3.073476623 &~~
3.336533166 &~~ 3.340286437 &~~ 3.343298255 \\ [0.5ex]\hline
$$
&~~ $$ &~~ $$ &~~ $B= B_{cr}$ &~~ $$ &~~
$$ &~~
$$ \\ [0.5ex]\hline

$\beta$ &~~ $3.84$ &~~ $1.21 $  &~~ $8.59\times10^{-1}$
&~~ $3.84\times10^{-1}$ &~~ $2.72 \times10^{-1}$ &~~ $1.21 \times10^{-1}$\\
$b_{cr}$ &~~ 3.977197683 &~~ 3.875208802  &~~ 3.794570716 &~~
3.605006776 &~~ 3.550994683 &~~ 3.511239776 \\ [0.5ex]\hline
$$
&~~ $$ &~~ $$ &~~ $B=10^{-1} B_{cr}$ &~~ $$ &~~
$$ &~~
$$ \\ [0.5ex]\hline

$\beta$ &~~ $3.84\times10^{-1}$ &~~ $1.21 \times10^{-1}$  &~~
$8.59\times10^{-2}$
&~~ $3.84\times10^{-2}$ &~~ $2.72 \times10^{-2}$ &~~ $1.21 \times10^{-2}$\\
$b_{cr}$ &~~ 3.986196724 &~~ 3.884373267  &~~ 3.803082454 &~~
3.609096180 &~~ 3.555556457 &~~ 3.514012326 \\ [0.5ex]\hline
$$
&~~ $$ &~~ $$ &~~ $B=10^{-2} B_{cr}$ &~~ $$ &~~
$$ &~~
$$ \\ [0.5ex]\hline

$\beta$ &~~ $3.84\times10^{-2}$ &~~ $1.21 \times10^{-2}$  &~~
$8.59\times10^{-3}$
&~~ $3.84\times10^{-3}$ &~~ $2.72 \times10^{-3}$ &~~ $1.21 \times10^{-3}$\\
$b_{cr}$ &~~ 3.986369643 &~~ 3.884665635  &~~ 3.803389832 &~~
3.609283622 &~~ 3.555759463 &~~ 3.514150243 \\ [0.5ex]\hline
$$
&~~ $$ &~~ $$ &~~ $B=10^{-4} B_{cr}$ &~~ $$ &~~
$$ &~~
$$ \\ [0.5ex]\hline

$\beta$ &~~ $3.84\times10^{-4}$ &~~ $1.21 \times10^{-4}$  &~~
$8.59\times10^{-5}$
&~~ $3.84\times10^{-5}$ &~~ $2.72 \times10^{-5}$ &~~ $1.21 \times10^{-5}$\\
$b_{cr}$ &~~ 3.986380595 &~~ 3.884690896  &~~ 3.803417676 &~~
3.609301547 &~~ 3.555778977 &~~ 3.514163865 \\ [0.5ex]\hline
$$
&~~ $$ &~~ $$ &~~ $B=10^{-6} B_{cr}$ &~~ $$ &~~
$$ &~~
$$ \\ [0.5ex]\hline

$\beta$ &~~ $3.84\times10^{-6}$ &~~ $1.21 \times10^{-6}$  &~~
$8.59\times10^{-7}$
&~~ $3.84\times10^{-7}$ &~~ $2.72 \times10^{-7}$ &~~ $1.21 \times10^{-7}$\\
$b_{cr}$ &~~ 3.986380696 &~~ 3.884691115  &~~ 3.803417948 &~~
3.609301957 &~~ 3.555779169 &~~ 3.514163998

\\ [1ex] \hline\hline
\end{tabular}
\end{table*}
%
%
%
%
%
%
\begin{table}[ht]
\caption{\label{table4} The value of parameter $\beta_{cr}$ at
which we have $b_{cr}=3$ and magnetic field starts acting as a
cosmic censor is tabulated here. The ratio of magnetic field to
the critical value is also specified. We find that the magnetic 
field is around $10$ times larger than the critical magnetic field 
$B_{cr}$ when it starts acting as a cosmic censor. } \centering
\begin{tabular}{c c c c c c c }
\hline\hline $$ & $$ & $$ & $b_{cr}=3$ & $$ &
$$
\\
[0.5ex]\hline

$\e$ & $0.01$ & $0.01$ & $0.01$ & $0.01$ & $0.01$  \\
[0.5ex]
$m$ & 0.0001 & 0.0001  & 0.0001 &
0.0001 & 0.0001 \\$\tilde{c}$& $10^{-3}$ & $10^{-4}$ & $5\times 10^{-5}$ & $10^{-5}$ & $5\times10^{-6}$\\
[0.5ex]\hline$\beta_{cr}$ & $37.0809$ & $12.3265$ & $9.1618$ &
$5.4369$ & $4.7714$ \\$$ & $-36.8021$ & $-12.3011$ & $-9.1490$ &
$-5.4342$ & $-4.7699$ \\ [0.5ex]\hline

$B/B_{cr}$ & $9.6514$ & $10.1457$ & $10.6644$ & $14.1512$ &
$17.5631$
\\& $9.5788$ & $10.1248$ & $10.6495$ & $14.1442$ &
$17.5576$ \\ [1ex] \hline\hline
\end{tabular}
\end{table}
%
We now find the critical value of the magnetic field for which the square root of change in the
Kretschmann scalar between the extremal and near-extremal configurations at horizon is equal to the
trace of the energy momentum tensor of the magnetic field at the horizon.
\begin{eqnarray}
K=(K_{1}-K_{2})^{1/2}\sim T=g_{\mu\nu}T^{\mu\nu}, \,
\end{eqnarray}
We set $x=\cos\theta=0$ as we restrict ourselves to the equatorial plane.

For the value chosen in this paper $\epsilon=0.01$, the critical value of the magnetic field is given by
\begin{eqnarray}
B_{cr}\sim0.8591.
\end{eqnarray}
%

We have shown that the magnetic field can possibly act as a cosmic
censor preventing a charged test particle from entering the black
hole that can turn it into a naked singularity. We now compare the strength 
of the requisite magnetic field by comparing its backreaction with that 
of the test particle. That can be done by comparing strength of the magnetic field 
with that of the critical magnetic field defined above $B_{cr}$.

We stated the values of parameters $\beta$ and $\tilde{c}$ while
calculating the effective potential in the previous sections to
understand whether magnetic field can prevent charged particle
from entering the black hole. In order to compare the backreaction
of magnetic field with that of test particle we have to calculate
the value of the magnetic field and compare it with the critical
value.
%
%
Eliminating the charge $e$ in the expression for the parameter
$\beta$ one can get
\begin{eqnarray}\label{eq.beta}
\beta=\sqrt{\frac{2\tilde{c}}{m}}M B .
\end{eqnarray}
This equation allows us to relate the magnetic field $B$ with
$\beta$ and $\tilde{c}$ for the given mass of the test particle
$m$. We may take $M=1$. Since we assume that $\tilde{\d E}\sim2\d
E/m$ is large and $\d E \sim \epsilon$, we take
$m\sim\epsilon^{2}$.

We now analyze the effective potential for the same values of
parameters we have chosen in the previous section and see whether
magnetic field can act as a cosmic censor by calculating the
critical value $b_{cr}$. We compute and compare the magnetic field
with the critical value.

\begin{figure*}
\centering
%
a)  \includegraphics[width=0.45\textwidth]{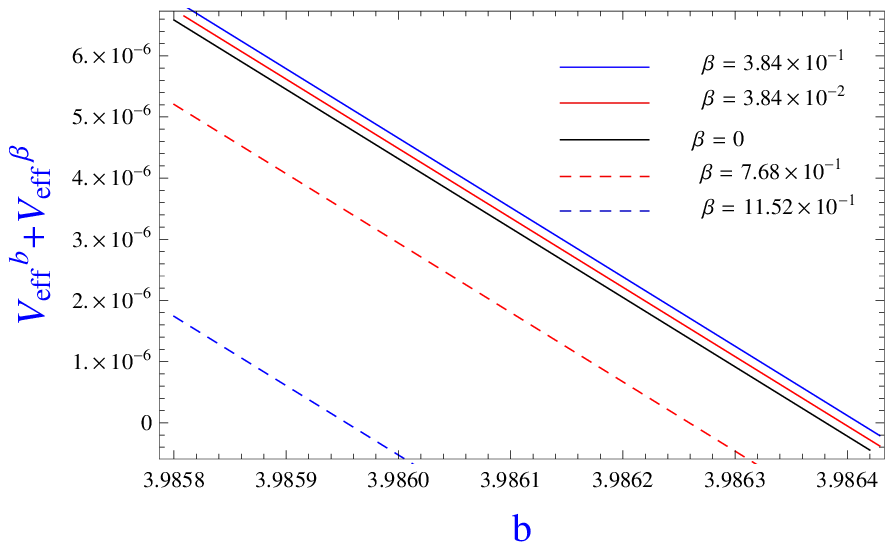}
b)  \includegraphics[width=0.45\textwidth]{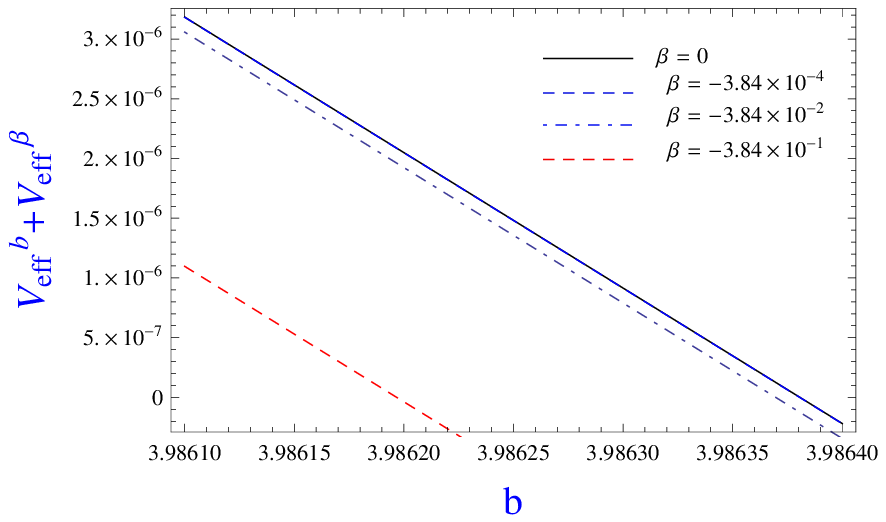}
\caption{\label{weakMF} The value of the effective potential at
maximum as a function of $b$ is plotted for positive as well as
negative values of $\beta$ and for charge parameter
$\tilde{c}=10^{-3}$. The magnetic field is sufficiently smaller
than the critical value. $b_{cr}$ initially increases and then it
decreases. We have $b_{cr}>3$. Thus when bacreaction of the
magnetic field can be ignored it does not serve as a cosmic
censor.}
\end{figure*}

We have tabulated the values of $b_{cr}$ for different values of
$\beta>0$ in Table~\ref{table2} and for $\beta<0$ in Table
\ref{table3}. For the values of $\beta$ where magnetic field is
sufficiently small as compared to the critical value, we find that
$b_{cr}$ initially increases and then it tends to decrease as we
increase parameter $\beta$. As long as magnetic field is smaller
as compared to the critical value, we get $b_{cr}>3$ and thus 
there is an allowed range of values of $b$ for
which it is possible for charged particle to enter the black hole
and turn it into a naked singularity. For the values of $\beta$
for which there is no allowed range of $b$, we find that the
magnetic field is slightly larger than the critical value. In
Table ~\ref{table4} we have tabulated the values of $\beta$ beyond
which magnetic field starts acting as a cosmic censor and the
ratio $B/B_{cr}$. Thus the backreaction of magnetic field on the
background spacetime is comparable to the tiny backreaction 
of the test particle. Thus the magnetic field necessary to restore
the cosmic censorship is extremely small.

However since the backreaction of the magnetic field on the background spacetime 
is comparable to that of the test particle one must take into account the 
effect of magnetic field on the background spacetime while analyzing whether or not 
it can act as a cosmic censor. Various exact magnetized black hole solutions
have been obtained \cite{Ernst,Gibbons}. However there is no exact solution that represents 
near-extremal black hole. Thus it is a daunting task to take into account the effect of 
magnetic field on the background metric. However as it was shown in \cite{Gibbons} in the
context of the magnetized Reissner-Nordstr\"{o}m black hole that 
the small magnetic field does not change either the horizon radius or the extremality condition.
Similar results are also expected to hold good in the context of the near extremal rotating 
black hole. It seems that the effect of the magnetic field on the background metric
does not assist the process of destroying black holes. Thus the conclusions obtained 
by ignoring the backreaction are expected to hold good even when the backreaction is taken 
into account.

\section{Conclusions}
\label{Sec:Conclusion}

In this paper we studied the process of turning black hole into a
naked singularity by throwing in the test particle with
appropriate values of the geodesic parameters. It is possible to
turn a near extremal Kerr black hole into a Kerr-Newmann naked
singularity using a charged test particle. Typically in the
astrophysical context black holes are surrounded with a magnetic
field which would exert a Lorentz force on the charged particle
affecting its motion. Thus we study the effect of the test
magnetic field on the process of destroying black hole. We invoke
a weak magnetic field which takes a constant value at infinity and
is aligned with the axis of symmetry of Kerr geometry. We show
that for a sufficiently large values of magnetic field it is not
possible for a particle with the appropriate values of geodesic
parameters to enter the black hole and turn it into the naked
singularity. Thus it appears that test magnetic field could serve
as a cosmic censor. To gauge the strength of the requisite magnetic field 
we compute trace of its energy momentum tensor 
and compare it with the square root of the change in the Kretschmann scalar at
the horizon between extremal and near-extremal configurations
which is a measure of the effect of the test particle on the
background spacetime. We find that when the magnetic field acts as a
cosmic censor its backreaction is slightly larger than that of the test 
particle. Therefore we need an extremely small magnetic field to restore 
the cosmic censorship in the process of destroying near extremal
Kerr black hole with a charged test particle. However since the backreaction 
of the magnetic field is stronger than that of the test particle one must take account 
its effect on the metric which is difficult to implement in the absence of 
any near extremal rotating magnetized black hole solution. But since, 
neither the horizon radius nor the extremality condition for the magnetized 
Reissner-Nordstr\"{o}m black hole is affected due to the small magnetic field, 
it seems that the magnetic field does not assist the process of destroying 
black hole. Therefore the results obtained without considering the backreaction 
of the magnetic field are expected to hold good even in the presence of the 
backreaction and magnetic field would act as a cosmic censor. 

\section*{Acknowledgments}
SS and BA thank TIFR and IUCAA for the warm hospitality during
their stay in Mumbai and Pune, India. This research is supported
in part by Projects No. F2-FA-F113, No. EF2-FA-0-12477, and No.
F2-FA-F029 of the UzAS and by the ICTP through the OEA-PRJ-29 and
the OEA-NET-76 projects and by the Volkswagen Stiftung (Grant No.
86 866).

 \edoc